\newfam\msbfam
\font\twlmsb=msbm10 at 12pt
\font\eightmsb=msbm10 at 8pt
\font\sixmsb=msbm10 at 6pt
\textfont\msbfam=\twlmsb
\scriptfont\msbfam=\eightmsb
\scriptscriptfont\msbfam=\sixmsb
\def\cj{\fam\msbfam}

\def\C{{\cj C}}

\def\R{{\cj R}}

\def\Z{{\cj Z}}

\centerline{\bf GEOMETRY OF THE AHARONOV-BOHM EFFECT}

\

\centerline{\bf R. S. Huerfano$^{1,2}$, M. A. L\'opez$^2$ and M. Socolovsky$^2$} 

\

\centerline{\it $^1$ Departamento de Matem\'aticas, Universidad Nacional de Colombia, Bogot\'a, Colombia}

\centerline{\it $^2$ Instituto de Ciencias Nucleares, Universidad Nacional Aut\'onoma de M\'exico}
\centerline{\it Circuito Exterior, Ciudad Universitaria, 04510, M\'exico D. F., M\'exico} 

\

{\it We show that the connection responsible for any abelian or non abelian Aharonov-Bohm effect with $n$ parallel ``magnetic'' flux lines in $\R^3$, lies in a trivial $G$-principal bundle $P\to M$, i.e. $P$ is isomorphic to the product $M\times G$, where $G$ is any path connected topological group; in particular a connected Lie group. We also show that two other bundles are involved: the universal covering space $\tilde{M}\to M$, where path integrals are computed, and the associated bundle $P\times_G \C^m \to M$, where the wave function and its covariant derivative are sections.}

\


\

As is well known, the magnetic Aharonov-Bohm (A-B) effect $^{1,2}$ is a gauge invariant, non local quantum phenomenon, with gauge group $U(1)$, which takes place in a non simply connected space. It involves a magnetic field in a region where an electrically charged particle obeying the Schroedinger equation cannot enter, i.e. the ordinary 3-dimensional space minus the space occupied by the solenoid producing the field; in the ideal mathematical limit, the solenoid is replaced by a flux line. Locally, the particle couples to the magnetic potential $\vec{A}$ but not to the magnetic field $\vec{B}$; however, the effect is gauge invariant since it only depends on the flux of $\vec{B}$ inside the solenoid. 

\

The fibre bundle theoretic description of this kind of phenomena has proved to be very useful to obtain a more profound insight into the relation between physical processes and pure mathematics. $^3$ In the present case, since by symmetry, the dimension along the flux line can be ignored, the problem reduces to the effect on the charged particle of an abelian connection in a $U(1)$-bundle with base space the plane minus a point. The wave function representing the particle is a section of an associated vector bundle. As shown in ref. 4, the bundle turns out to be trivial and then its total space is isomorphic to the product $(\R^2-\{point\})\times U(1)$. Since $\R^2$ is topologically equivalent to an open disk, and $U(1)$ is the unit circle, the bundle structure is summarized by $$U(1)\to {T}^2_{\circ*}\to {D}^2_{\circ*} \eqno{(1)}$$ where $D^2_{\circ*}$ is the open disk minus a point and $T^2_{\circ*}$ is the open solid 2-torus minus a circle. 

\

As suggested by Wu and Yang $^6$, Yang-Mills fields can give rise to non abelian A-B effects. In ref. 6 the authors studied an $SU(2)$ gauge configuration leading to an A-B effect; later, several authors studied the effect with gauge groups $SU(3)$ $^7$ and $U(N)$ $^8$. Also, in refs. 9-14 the effect was studied in the context of gravitation theory. In all these examples, as in the  magnetic case, there is a principal bundle structure $$\xi: G\to P\buildrel {\pi_G}\over\longrightarrow M \eqno{(2)}$$ whose total space $P$, where the connection giving rise to the effect lies, is however never specified. We shall restrict ourselves to connections $\omega$ giving rise to $n$ ($n=1,2,3,...$) flux lines in $\R^3$. In this note we prove the following: 

\

{\it Theorem}: Let $G$ be a path connected topological group (for example a connected Lie group) and $\xi$ a continuous principal $G$-bundle over $\R^3 - \{n \ parallel \ lines\}$, $n=1,2,3,...$ Then the bundle $\xi$ is trivial, i.e. isomorphic to the product bundle.

\

{\it Proof of the theorem}

\

The classification of bundles over $\R^3 \backslash \{n$ parallel lines$\}$ is the same as that over $\R^2 \backslash \{n$ points$\}$ which is topologically equivalent to $D^2_\circ \backslash \{n$ points $\}\equiv D^2_{\circ * n}$. Denote this set of points by $\{b_1,...,b_n\}$. By symmetry along the dimension of the flux tubes, $D^2_{\circ*n}$ is the space where it can be considered that the charged particles move. 

Let $x_0$ be a point in $D^2_{\circ*n}$. We construct a bouquet of $n$ loops $\gamma_1$, $\gamma_2$,..., $\gamma_n$ through $x_0$, with the k-th loop sorrounding the point $b_k$, $k=1,2,...,n$. This space is homeomorphic to the wedge product (or reduced join) $S^1\vee...\vee S^1\equiv \vee_n S^1\equiv S^1_{(1)}\vee...\vee S^1_{(n)}$ of $n$ circles $^{15}$, and the classification of bundles over $D^2_{\circ*n}$ is the same as that over $\vee_n S^1$, namely $${\cal B}_{D^2_{\circ*n}}(G)={\cal B}_{\vee_n S^1}(G) \eqno{(3)}$$ where ${\cal B}_M(G)$ is the set of isomorphism classes of $G$-bundles over $M$ $^{16}$. 

By explicit construction we shall prove that, up to isomorphism, the unique $G$-bundle over $\vee_n S^1$ is the product bundle $\vee_n S^1\times G$ (which is a purely topological result). With this aim, we cover the circle $S_{(k)}^1$ with two open sets $U_{k+}$ and $U_{k-}$ such that $$U_{k+}\cap U_{k-}\simeq\{x_0,a_k\}, \ k=1,...,n, \eqno{(4)}$$ $$U_{i+}\cap U_{j+}\simeq U_{i-}\cap U_{j-}\simeq U_{i+}\cap U_{j-}\simeq \{x_0\}, \ i,j=1,...,n, \ i\neq j \eqno{(5)}$$ where $\simeq$ denotes homotopy equivalence, and $a_k\in S^1_{(k)}$ with $a_k \neq x_0$. We then have $$2\times\pmatrix{2n \cr 2 \cr}+2n={{2\times(2n)!}\over {(2n-2)!2!}}+2n=4n^2 \eqno{(6)}$$ transition functions $$g_{\alpha, \beta}:U_\alpha \cap U_\beta \to G \eqno{(7)}$$ with $g_{\beta,\alpha}=g_{\alpha,\beta}^{-1}$. Up to homotopy they are given by $$g_{k+,k-}:\{x_0,a_k\}\to G, \ x_0 \mapsto g_{0k}, \ a_k\mapsto g_k, \ k=1,...,n \eqno{(8)}$$ and $$g_{i+,j+},g_{i-,j-},g_{i+,j-}:\{x_0\}\to G, \ i,j=1,...,n, \ i\neq j,$$ $$x_0\mapsto g_{ij++}, \ x_0\mapsto g_{ij--}, \ x_0\mapsto g_{ij+-}. \eqno{(9)}$$ The $2n$ transition functions $g_{i+,i+}$, $g_{i-,i-}$, $i\in\{1,...,n\}$, give the identity in $G$. The number of cocycle relations $g_{\beta,\alpha}g_{\alpha,\gamma}=g_{\beta,\gamma}$ on the $2n(2n-1)$ non trivial transition functions $g_{\mu,\nu}$ is $\pmatrix{2n \cr 3\cr}$=${{n(2n-1)(2n-2)}\over{3}}$.

Let $g^\prime_{k+,k-}:\{x_0,a_k\}\to G$ be given by $g^\prime_{k+,k-}(x_0)=g^\prime_{0k}$, $g^\prime_{k+,k-}(a_k)=g^\prime_k$. Since $G$ is path connected, there exist continuous paths $$c^{(k)}_0:[0,1]\to G, \ t\mapsto c_0^{(k)}(t) \ \ with \ \ c_0^{(k)}(0)=g_{0k}, \ c_0^{(k)}(1)=g^\prime_{0k}$$ and $$c_k^{(k)}:[0,1]\to G, \ t\mapsto c_k^{(k)}(t) \ \ with \ \ c_k^{(k)}(0)=g_k, \ c_k^{(k)}(1)=g^\prime_k.$$ Then the continuous function $$H_k:\{x_0,a_k\}\times [0,1]\to G$$ given by $$H_k(x_0,t)=c_0^{(k)}(t), \ \ H_k(a_k,t)=c_k^{(k)}(t) \eqno{(10)}$$ is a homotopy between $g_{k+,k-}$ and $g^\prime_{k+,k-}$ since $$H_k(x_0,0)=g_{0k} \ \ and \ \  H_k(a_k,0)=g_k$$ i.e. $$H_k\vert_{\{x_0,a_k\}\times \{0\}}=g_{k+,k-}$$ and $$H_k(x_0,1)=g^\prime_{0k} \ \ and \ \ H_k(a_k,1)=g^\prime_k$$ i.e. $$H_k\vert_{\{x_0,a_k\}\times\{1\}}=g^\prime_{k+,k-}.$$ Then the homotopy class of maps from $\{x_0,a_k\}$ to $G$ has only one element, namely $[g_{k+,k-}]_{\sim}$ i.e. $$[\{x_0,a_k\},G]_{\sim}=\{[g_{k+,k-}]_{\sim}\}. \eqno{(11)}$$

Similarly, let $g_{i\mu,j\nu}$ with $\mu,\nu \in\{+,-\}$ be any of the functions in (9), and let $g^\prime_{i\mu,j\nu}:\{x_0\}\to G$ be given by $$g^\prime_{i\mu,j\nu}(x_0)=g^\prime_{ij,\mu\nu}; \eqno{(12)}$$ then the map $$H_{i\mu,j\nu}:\{x_0\}\times[0,1]\to G, \ \ H_{i\mu,j\nu}(x_0,t)=c_{i\mu,j\nu}(t) \eqno{(13)}$$ with $c_{i\mu,j\nu}:[0,1]\to G$ a continuous path in $G$ satisfying $c_{i\mu,j\nu}(0)=g_{ij\mu\nu}$ and $c_{i\mu,j\nu}(1)=g^\prime_{ij\mu\nu}$, is a homotopy between $g_{i\mu,j\nu}$ and $g^\prime_{i\mu,j\nu}$ i.e. $$[\{x_0\},G]_{\sim}=\{[g_{i\mu,j\nu}]_{\sim} \}. \eqno{(14)}$$ It is then easy to show that, if we define (constant) functions $$\Lambda_{k\mu}:U_{k\mu}\to G, \ \ p \to \lambda_{k\mu}, \ \ k=1,...,n, \ \ \mu=+,-,\eqno{(15)}$$ then $$g^\prime_{k+,k-}(x_0)=\lambda_{k-}g_{k+,k-}(x_0)\lambda_{k+}^{-1}, \eqno{(16)}$$ $$g^\prime_{k+,k-}(a_k)=\lambda_{k-}g_{k+,k-}(a_k)\lambda_{k+}^{-1}, \eqno{(17)}$$ for $k=1,...,n$, and $$g^\prime_{i\mu,j\nu}(x_0)=\lambda_{j\nu}g_{i\mu,j\nu}(x_0)\lambda_{i\mu}^{-1} \eqno{(18)}$$ for $i,j\in\{1,...,n\}$, $i\neq j$, $\mu,\nu\in\{+,-\}$. In fact, for arbitrary two sets of homotopic transition functions $g_{\beta,\alpha}^\prime \sim g_{\beta,\alpha}$, one has continuous maps $H:(U_\beta \cap U_\alpha)\times [0,1]\to G$ such that $H(x,0)=g_{\beta,\alpha}(x)$ and $H(x,1)=g_{\beta,\alpha}^\prime(x)$. Given a map $\Lambda_\alpha:U_\beta \cap U_\alpha \to G$, $x\to \Lambda(x)=\lambda_\alpha$, one defines $\bar{H}(x,1)=H(x,1)\Lambda_\alpha(x)$ and therefore $\bar{H}(x,1)=g_{\beta,\alpha}^\prime(x)\Lambda_{\alpha}(x)$. Now, one defines $\Lambda_\beta:U_\beta \cap U_\alpha \to G$ through $\Lambda_{\beta}(x)=\bar{H}(x,1)H(x,0)^{-1}=g_{\beta,\alpha}^\prime(x)\Lambda_{\alpha}(x)g_{\beta,\alpha}(x)^{-1}$ i.e. $g_{\beta,\alpha}^\prime(x)=\Lambda_\beta(x)g_{\beta,\alpha}(x)\Lambda_\alpha(x)^{-1}$ (*) for all $x\in U_\beta\cap U_\alpha$. In our case, for $U_{k+}\cap U_{k-}\simeq \{x_0,a_k\}$, the formulae corresponding to (*) are $$g_{k+,k-}^\prime(x_0)=\Lambda_{k-}(x_0)g_{k+,k-}(x_0)\Lambda_{k+}(x_0)^{-1}$$ and $$g_{k+,k-}^\prime(a_k)=\Lambda_{k-}(a_k)g_{k+,k-}(a_k)\Lambda_{k+}(a_k)^{-1}.$$ Extending continuously (as constants) $\Lambda_{k+}$ and $\Lambda_{k-}$ respectively to the open sets $U_{k+}$ and $U_{k-}$: $$\Lambda_{k+}:U_{k+}\to G, \ p\mapsto \lambda_{k+}, \ \ \Lambda_{k-}:U_{k-}\to G, \ q\mapsto \lambda_{k-}$$ (in particular one has $\Lambda_{k+}(x_0)=\Lambda_{k+}(a_k)=\lambda_{k+}$ and $\Lambda_{k-}(x_0)=\Lambda_{k-}(a_k)=\lambda_{k-}$), one obtains equations (16) and (17). Proceeding similarly for the cases $U_{i\mu}\cap U_{j\nu}\simeq\{x_0\}$, one gets equation (18).

Then, up to isomorphisms, and according to a general theorem for coordinate bundles $^{17}$, there is a unique $G$-bundle generated by the transition functions given by the equations (8) and (9), namely the product bundle. We then have $${\cal B}_{\vee_n S^1}(G)=\{[G\to\vee_n S^1 \times G\to \vee_n S^1]\}, \eqno{(19)}$$ where $[ \ \ \ ]$ denotes here the equivalence class of bundles isomorphic to the product bundle.    QED 

\

For the cases of the examples in refs. 1, 6 and 7, we have, respectively, $${\cal B}_{D^2_{\circ *}}(U(1))=\{[U(1)\to D^2_{\circ *}\times U(1)\to D^2_{\circ *}]\}, \eqno{(20)}$$ $${\cal B}_{D^2_{\circ *}}(SU(2))=\{[SU(2)\to D^2_{\circ *}\times SU(2)\to D^2_{\circ *}]\}, \eqno{(21)}$$ and $${\cal B}_{D^2_{\circ *2}}(SU(3))=\{[SU(3)\to D^2_{\circ *2}\times SU(3)\to D^2_{\circ *2}]\}. \eqno{(22)}$$ 

\

In the gravitational case, the gauge group is $SL(2,\C)$ $^{18}$, which is the universal covering group of the connected component of the Lorentz group $L^\uparrow_+$; then, for weak gravitational fields with a distribution of $n$ gravitomagnetic flux lines as above we would have the trivial bundle $$SL(2,\C)\to D^2_{\circ *n}\times SL(2,\C)\to D^2_{\circ *n}. \eqno{(23)}$$

\

We want to stress that, even if the A-B connection is flat (though not exact), one of the sufficient conditions for the automatic triviality of the A-B bundle fails: though paracompact, the bouquet $\vee_n S^1$ is not simply connected. (See corollary 9.2. in ref. 21, p. 92.) In addition, there is not a priori any physical reason why, under the specified conditions on the ``magnetic flux'', the corresponding A-B  bundle should be trivial.

\

It is interesting to notice that there are two other fibre bundles related to the A-B effect. The first bundle is the {\it universal covering space} $^{19}$ of the base manifold (``laboratory'' or physical space where the particles coupled to the A-B potential move) which is the $\pi_1(M;x_0)$-(non trivial) bundle $\xi_c:\tilde{M}\buildrel {\pi}\over\longrightarrow M$, where $\pi_1(M;x_0)$ ($\equiv \pi_1(M)$ if $M$ is connected) is the fundamental group of $M$. In our case, $$\pi_1(\vee_n S^1;x_0)\cong \pi_1(\R^2\backslash \{n \ points\};x_0)\cong <\{c_1,...,c_n\}>$$ is the freely generated group with $n$ non commuting generators $c_1,...c_n$. In particular, for the original abelian A-B effect with group $U(1)$, $\tilde{\R^{2*}}=RS(Log)$: the Riemann surface of the logarithm, and $\pi_1(\R^{2*})\cong \Z$. 

\

The {\it particle propagator} in $M$, $K(x^{\prime\prime},t^{\prime\prime}; x^\prime,t^\prime)$ with $t^{\prime\prime}>t^\prime$, is a sum of homotopy propagators $^7$ multiplied by corresponding gauge factors $^{20}$: the former are given by unrestricted path integrals computed in $\tilde{M}$, the paths in these path integrals project onto the corresponding homotopy classes of paths in the non simply connected space $M$; the latter are Wilson loops given by $$Texp\int_{\pi(c)}\vec{A}\cdot d\vec{l}$$ where $T$ denotes time order, $\vec{A}$ is the A-B potential, and $c$ is a loop in $\tilde{M}$ beginning  and ending respectively at $y_0$ and $y^{\prime\prime}$ in $\pi^{-1}(\{x^{\prime\prime}\})\buildrel{\Psi}\over\cong \pi_1(M)$, with $y_0$ fixed and arbitrary. Then one has the group homomorphism (many-to-one or one-to-one) $$y^{\prime\prime}\buildrel {\Psi}\over \longrightarrow \Psi(y^{\prime\prime})\buildrel{\varphi}\over\longrightarrow Texp\int_{\pi(c)}\vec{A}\cdot d\vec{l} \ \ \in G \eqno{(24)}$$ whose image in $G$, $\varphi(\pi_1(M))$, responsible for the A-B effect, is the {\it holonomy} of the connection. $^{21}$

\

The second bundle is the {\it associated complex vector bundle} $\xi_{\C ^m}: \C ^m - P_{\C ^m}\buildrel {\pi_{\C ^m}}\over \longrightarrow M$ ($m=2s+1$ is the dimension of the spinor space and $s$ is the spin; for scalar particles $m=1$), where $$P_{\C^m}=(M\times G)\times_G \C^m=\{[((x,g),\vec{z})]\}_{((x,g),\vec{z})\in(M\times G)\times_G\C^m},$$ $$[((x,g),\vec{z})]=\{((x,gg^\prime),g^{\prime -1}\vec{z})\}_{g^\prime \in G}. \eqno{(25)}$$ $\xi_{\C^m}$ is trivial since $\xi$ is trivial, and the quantum mechanical {\it wave functions} of the particles are global sections of $\xi_{\C^m}$: $$\psi\in \Gamma(\xi_{\C^m})$$ i.e. $\psi:M\to P_{\C^m}$ with $\pi_{\C^m}\circ \psi= Id_M$. 

\

Notice that while the propagator is computed in $\xi_c$, the wave function lies in $\xi_{\C^m}$, with $$\psi(x^{\prime\prime},t^{\prime\prime})=\int_M dx^\prime K(x^{\prime\prime},t^{\prime\prime};x^\prime,t^\prime)\psi(x^\prime,t^\prime).\eqno{(26)}$$ If $\omega$ is the A-B connection in $P$, then the coupling $\omega-\psi$ is the covariant derivative $$\nabla_V^\omega \psi=\psi_{V^\uparrow(\gamma_\psi)}\in \Gamma(\xi_{\C^m}), \eqno{(27)}$$ where $V$ is a vector field in $M$, $V^\uparrow$ its horizontal lifting in $P$ by $\omega$, $\gamma_\psi$ and $V^\uparrow(\gamma_\psi)$ are equivariant functions from $P$ to $\C^m$ with $\gamma_\psi(p)=\vec{z}$  where $\psi(\pi_G(p))=[p,\vec{z}]$, and $\psi_{V^\uparrow(\gamma_\psi)}(x)=[p,V^\uparrow(\gamma_\psi)(p)]$ for any $p\in \pi_G^{-1}(\{x\})$. Locally, of course, $\nabla_V^\omega\psi$ reproduces the usual minimal coupling between $\vec{A}$ and $\psi$.

\

Finally, though $\varphi:\pi_1(M)\to G$ is a group homomorphism, and $\tilde{M}\buildrel\ {f}\over \longrightarrow M \times G$ given by $f(y)=(\pi(y),1)$ is a canonical map, there is no bundle map between $\xi_c$ and $\xi$: the pair of functions $(f\times\varphi,f)$ is {\it not} a principal bundle homomorphism.   

\

In summary, the three bundles are related by the following diagram:

$$\matrix{\pi_1(M) & \buildrel{\varphi}\over\longrightarrow & G & & & &  \C^m \cr
          \downarrow & &  \downarrow & & & &  \vert \cr
          \tilde{M} & \buildrel{f}\over \longrightarrow & M\times G & &  \buildrel{\iota}\over \longrightarrow & &  (M\times G)\times_G\C^m \cr
          \downarrow \pi & & \downarrow \pi_G & & & & \downarrow \pi_{\C^m}\uparrow \psi \uparrow \nabla^\omega _V \psi \cr
          M & = & M & & = & & M \cr}$$ where $\iota$ is the canonical injection of the bundle $\xi$ into its associated bundle i.e. $\iota(p)=[p,0]$. 

\

{\bf Acknowledgments}

\

This work was partially supported by the grant PAPIIT-UNAM IN103505. M. S. thanks for hospitality to the Instituto de Astronom\'\i a y F\'\i sica del Espacio (UBA-CONICET, Argentina), and the University of Valencia, Spain, where part of this work was done. 

\

{\bf References}

\

1. Aharonov, Y. and Bohm, D.(1959). Significance of Electromagnetic Potentials in Quantum Theory, {\it Physical Review}, {\bf 115}, 485-491.

\

2. Chambers, R. G. (1960). Shift of an Electron Interference Pattern by Enclosed Magnetic Flux, {\it Physical Review Letters}, {\bf 5}, 3-5.

\

3. Daniel, M. and Viallet, C. M. (1980). The Geometrical Setting of Gauge Theories of the Yang-Mills Type, {\it Reviews of Modern Physics}, {\bf 52}, pp. 175-197.

\

4. Aguilar, M. A. and Socolovsky, M. (2002). Aharonov-Bohm Effect, Flat Connections, and Green's Theorem, {\it International Journal of Theoretical Physics}, {\bf 41}, 839-860. (For a review see, e.g. ref. 5.)

\

5. Socolovsky, M. (2006). Aharonov-Bohm Effect, in {\it Encyclopedia of Mathematical Physics}, Elsevier, Amsterdam, pp. 191-198.

\

6. Wu, T. T. and Yang, C. N. (1975). Concept of non Integrable Phase Factors and Global Formulation of Gauge Fields, {\it Physical Review D}, {\bf 12}, 3845-3857.

\

7. Sundrum, R. and Tassie, L. J. (1986). Non-Abelian Aharonov-Bohm Effects, Feynman Paths, and Topology, {\it Journal of Mathematical Physics}, {\bf 27}, 1566-1570.

\

8. Botelho, L. C. L. and de Mello, J. C. (1987). A Non-Abelian Aharonov-Bohm Effect in the Framework of Feynman Pseudoclassical Path Integrals, {\it Journal of Physics A: Math. Gen.}, {\bf 20}, 2217-2219. 

\

9. Harris, E. G. (1996). The gravitational Aharonov-Bohm effect with photons, {\it American Journal of Physics}, {\bf 64}, 378-383.

\

10. Zeilinger, A., Horne, M. A. and Shull, C. G. (1983). Search for unorthodox phenomena by neutron interference experiments, {\it Proceedings International Symposium of Foundations of Quantum Mechanics}, Tokio, pp. 289-293.

\

11. Ho, Vu B. and Morgan, M. J. (1994). An experiment to test the gravitational Aharonov-Bohm effect, {\it Australian Journal of Physics}, {\bf 47}, 245-252.

\

12. Bezerra, V. B. (1987). Gravitational analogue of the Aharonov-Bohm effect in four and three dimensions, {\it Physical Review D}, {\bf 35}, 2031-2033.

\

13. Wisnivesky, D. and Aharonov, Y. (1967). Nonlocal effects in classical and quantum theories, {\it Annals of Physics}, {\bf 45}, 479-492.

\

14. Corichi, A. and Pierri, M. (1995). Gravity and Geometric Phases, {\it Physical Review D}, {\bf 51}, 5870-5875.

\

15. Greenberg, M. J. and Harper, J. R. (1981). {\it Algebraic Topology. A First Course}, Addison-Wesley, Redwood City, p. 126.  

\

16. Nash, C. and Sen, S. (1983). {\it Topology and Geometry for Physicists}, Academic Press, London, p. 262.

\

17. Idem 16., pp. 147-148.

\

18. Naber, G. L. (2000). {\it Topology, Geometry, and Gauge Fields, Interactions}, Springer-Verlag, New York, pp. 193-197.

\

19. Massey, W. S. (1991). {\it A Basic Course in Algebraic Topology}, GTM 56, Springer, New York, p. 132. 

\

20. Schulman, L. S. (1981). {\it Techniques and Applications of Path Integration}, Wiley, New York, pp. 205-207.  

\

21. Kobayashi, S and Nomizu, K. (1963). {\it Foundations of Differential Geometry, Vol. 1}, Wiley, New York, p. 71.

\

\

\

\

\

\

\

e-mails: 

\

rshuerfanob@unal.edu.co

\

alicia@nucleares.unam.mx

\

socolovs@nucleares.unam.mx

\end

\

2. {\bf Galilean group, its universal covering group, and spinors}

\

The connected component of the galilean group $G_0$ consists of the set of 4$\times$4 matrices $$g=\pmatrix{R&\vec{V}\cr 0&1\cr} \eqno{(6)}$$ with $R$ in the 3-dimensional rotation group $SO(3)$, boost velocity $\vec{V}$ in $\R^3$, composition law $$g_2g_1=\pmatrix{R_2 &\vec{V}_2 \cr 0 & 1 \cr}\pmatrix{R_1 & \vec{V}_1 \cr 0 &1\cr}=\pmatrix{R_2R_1 & \vec{V}_2+R_2\vec{V}_1 \cr 0 & 1 \cr}, \eqno{(6a)}$$ identity $$\pmatrix{I & 0 \cr 0 &1 \cr}, \ \ \ I=\pmatrix{1&0&0\cr 0&1&0 \cr 0&0&1}, \eqno{(6b)}$$ and inverse $$\pmatrix{R & \vec{V}\cr 0 &1}^{-1}=\pmatrix{R^{-1} & -R^{-1}\vec{V} \cr 0 & 1}. \eqno{(6c)}$$ $G_0$ is a non abelian, non compact, connected but non simply connected six dimensional Lie group; like the connected component of the Lorentz group, its topology is that of the cartesian product of the real projective space with ordinary 3-space {\it i.e.} of $\R P^3\times \R ^3$. The action of $G_0$ on spacetime is given by $$G_0\times \R ^4\to \R ^4, \ (g,\pmatrix{\vec{x}^\prime \cr t^\prime \cr})\mapsto \pmatrix{\vec{x}\cr t \cr}=g\pmatrix{\vec{x}^\prime \cr t^\prime \cr}=\pmatrix{R\vec{x}^\prime+\vec{V}t^\prime \cr t^\prime}. \eqno{(7)}$$ Since one has the action $$\mu:SO(3)\times \R ^3 \to \R ^3, \ (R,\vec{x})\mapsto R\vec{x}, \eqno{(8)}$$then $G_0$ is isomorphic to the semidirect sum $\R ^3 \times_{\mu}SO(3)$: $\pmatrix{R & \vec{V} \cr 0 & 1 \cr}\mapsto (\vec{V},R)$ with composition law $$(\vec{V}^\prime,R^\prime)(\vec{V},R)=(\vec{V}^\prime+R^\prime\vec{V},R^\prime R). \eqno{(8a)}$$

\ 

The {\it universal covering group} of $G_0$ is given by the $\Z_2$-bundle $$\Z_2\to \hat{G}_0 \buildrel {\Pi}\over \longrightarrow G_0 \eqno{(9)}$$ where $$\hat{G}_0=\{\hat{g}=\pmatrix{T & \vec{V} \cr 0 & 1 \cr}, \ T\in SU(2), \ \vec{V}\in \R ^3\}, \eqno{(9a)}$$ and $\Pi$ is the $2\to 1$ group homomorphism $$\Pi(\hat{g})=\pmatrix{\pi(T) & \vec{V} \cr 0 & 1 \cr} \eqno{(9b)}$$ with $\pi:SU(2) \to SO(3)$ the well known projection $$\pi\pmatrix{z & w \cr -\bar{w} & \bar{z} \cr}=\pmatrix{Rez^2-Rew^2 & Imz^2+Imw^2 & -2Rezw \cr -Imz^2+Imw^2 & Rez^2+Rew^2 & 2Imzw \cr 2Rez\bar{w} & 2 Imz\bar{w} & \vert z\vert ^2 -\vert w \vert ^2 \cr}. \eqno{(9c)}$$ $\hat{G}_0$ is simply connected and has the topology of $S^3\times \R ^3$. Since $SU(2)$ acts on $\R ^3$: $$\hat{\mu}:SU(2)\times \R ^3\to \R ^3, \ (T,\vec{V})\mapsto \pi(T)\vec{V},\eqno{(10)}$$ one has the group isomorphism $$\hat{G}_0\ni\pmatrix{T & \vec{V} \cr 0 & 1 \cr}\mapsto (\vec{V},T) \in \R ^3 \times _{\hat{\mu}}SU(2); \eqno{(11)}$$ the composition law in $\hat{G}_0$ is given by $$\pmatrix{T^\prime & \vec{V}^\prime \cr 0 & 1 \cr}\pmatrix{T & \vec{V} \cr 0 & 1 \cr}= \pmatrix{T^\prime T & \vec{V}^\prime+\pi(T^\prime)\vec{V} \cr 0 & 1 \cr}, \eqno{(12)}$$ while the identity and inverse are respectively given by $$\pmatrix{I & 0 \cr 0 & 1 \cr}, \ I=\pmatrix{1 & 0 \cr 0 & 1 \cr} \eqno{(12a)}$$ and $$\pmatrix{T & \vec{V} \cr 0 & 1 \cr}^{-1}=\pmatrix{T^{-1} & -\pi(T^{-1})\vec{V} \cr 0 & 1 \cr}. \eqno{(12b)}
$$ 

\

Turning back to physics, for each mass value $m>0$, $\hat{G}_0$ acts on the infinite dimensional Hilbert space ${\cal L}^2_1$ of continuously differentiable and square integrable $\C ^2$-valued functions $\pmatrix{u \cr v \cr}$ on $\R ^4$, the Schroedinger-Pauli spinors. This action is defined as follows: $^5$ $$\hat{\mu}_m:\hat{G}_0\times {\cal L}^2_1 \to {\cal L}^2_1, \ (\pmatrix{T & \vec{V} \cr 0 & 1 \cr},\pmatrix{u \cr v \cr})\mapsto \pmatrix{T & \vec{V} \cr 1 & 0 \cr}\cdot \pmatrix{u & \cr v \cr}: \R ^4 \to \C ^2,$$ $$\pmatrix{\vec{x} \cr t \cr}\mapsto \pmatrix{T & \vec{V} \cr 0 & 1 \cr}\cdot \pmatrix{u \cr v \cr}\pmatrix{\vec{x} \cr t \cr}= e^{{{-im}\over {\hbar}}(\vec{V}\cdot\vec{x}+{{1}\over{2}}\vert \vec{V}\vert ^2 t)}T\pmatrix{u(\pi(T)\vec{x}+\vec{V}t,t) \cr v(\pi(T)\vec{x}+\vec{V}t,t)}. \eqno{(13)}$$ $\hat{\mu}_m$ is equivalent to the {\it representation} $$\tilde{\hat{\mu}}_m:\hat{G}_0 \to End({\cal L}^2_1), \ \tilde{\hat{\mu}}_m(\hat{g})(\pmatrix{u \cr v & \cr})=\hat{g}\cdot \pmatrix{u \cr v \cr}. \eqno{(13.a)}$$ At each $t$ one has the inner product $$(\pmatrix{u_2 \cr v_2 \cr},\pmatrix{u_1 \cr v_1 \cr})(t)=\int d^3\vec{x}(\bar{u}_2(\vec{x},t)u_1(\vec{x},t)+\bar{v}_2(\vec{x},t)v_1(\vec{x},t)) \eqno{(14a)}$$ and the norm $$\vert \vert \pmatrix{u \cr v \cr} \vert \vert ^2 (t)=(\pmatrix{u \cr v \cr}, \pmatrix{u \cr v \cr})(t)=\int d^3\vec{x}(\vert u(\vec{x},t)\vert ^2 + \vert v(\vec{x},t)\vert ^2). \eqno{(14b)}$$ The galilean transformation of the charge conjugate spinor $\psi_c$ is given by $$\psi_c \mapsto \bar{\hat{g}}\cdot\psi_c, \ \pmatrix{\bar{T} & \vec{V} \cr 0 & 1 \cr}\cdot \pmatrix{-\bar{v} \cr \bar{u} \cr}\pmatrix{\vec{x} \cr t}=e^{{{im}\over{\hbar}}(\vec{V}\cdot\vec{x}+{{1}\over{2}}\vert \vec{V} \vert ^2t)}\bar{T}\pmatrix{-\bar{v}(\pi(\bar{T})\vec{x}+\vec{V}t,t) \cr \bar{u}(\pi(\bar{T})\vec{x}+\vec{V}t,t)}. \eqno{(15)}$$ Finally, the galilean transformations of the electromagnetic potential $(\phi, \vec{A})$ and the magnetic field $\vec{B}$ are $$\phi(\vec{x},t)=\phi^\prime(\vec{x}^\prime, t^\prime), \ \vec{A}(\vec{x},t)=R\vec{A}^\prime(\vec{x}^\prime,t^\prime), \ \vec{B}(\vec{x},t)=R\vec{B}^\prime(\vec{x}^\prime,t^\prime) \eqno{(16)}$$ with $\vec{x}=R\vec{x}^\prime+\vec{V}t^\prime$ and $t=t^\prime$. 

\

{\it Remark}: Representations associated with different values of the mass are inequivalent. $^6$

\

3. {\bf Lagrangian formulation and galilean and gauge invariances}

\

The Pauli equations (1) and (5) can be formulated within the lagrangian framework. The lagrangian for equation (1) is $${\cal L}={{i\hbar}\over {2}}((({{\partial}\over{\partial t}}-{{iq}\over{\hbar}}\phi)\psi^\dagger) \psi-\psi^\dagger ({{\partial}\over{\partial t}}+{{iq}\over{\hbar}}\phi)\psi)+{{\hbar ^2}\over{2m}}(\nabla+{{iq}\over{\hbar c}}\vec{A})\psi^\dagger\cdot(\nabla-{{iq}\over{\hbar c}}\vec{A})\psi-{{q\hbar}\over{2mc}}\psi^\dagger\vec{\sigma}\cdot\vec{B}\psi$$ $$={{i\hbar}\over{2}}(\dot{\psi}^\dagger\psi-\psi^\dagger\dot{\psi})+{{\hbar ^2}\over{2m}}\nabla\psi^\dagger\cdot\nabla\psi+{{q^2}\over{2mc^2}}\psi^\dagger\vert \vec{A}\vert ^2\psi+{{i\hbar q}\over{2mc}}(\psi^\dagger\vec{A}\cdot\nabla\psi-\nabla\psi^\dagger\cdot\vec{A}\psi)-{{q\hbar}\over{2mc}}\psi^\dagger\vec{\sigma}\cdot\vec{B}\psi+q\psi^\dagger\phi\psi, \eqno{(17)}$$ and equation (1) amounts to the variational equation $${{\delta}\over{\delta\psi^\dagger(\vec{x},t)}}S=0 \eqno{(18)}$$ where $S$ is the action $$S=\int dt\int d^3\vec{x}{\cal L}(\vec{x},t). \eqno{(19)}$$ Under the charge conjugation operation $${\cal L}\to {\cal L}_c=K{\cal L}=-{{i\hbar}\over {2}}((({{\partial}\over{\partial t}}+{{iq}\over{\hbar}}\phi)\psi^\dagger_c)\psi_c-\psi^\dagger_c ({{\partial}\over{\partial t}}-{{iq}\over{\hbar}}\phi)\psi_c)+{{\hbar ^2}\over{2m}}(\nabla-{{iq}\over{\hbar c}}\vec{A})\psi^\dagger_c\cdot(\nabla+{{iq}\over{\hbar c}}\vec{A})\psi_c+{{q\hbar}\over{2mc}}\psi^\dagger_c\vec{\sigma}\cdot\vec{B}\psi_c$$ 
$$=-{{i\hbar}\over{2}}(\dot{\psi}^\dagger_c\psi_c-\psi_c^\dagger\dot{\psi}_c)+{{\hbar ^2}\over{2m}}\nabla\psi^\dagger_c\cdot\nabla\psi_c+{{q^2}\over{2mc^2}}\psi^\dagger_c\vert \vec{A}\vert ^2\psi_c-{{i\hbar q}\over{2mc}}(\psi_c^\dagger\vec{A}\cdot\nabla\psi_c-\nabla\psi^\dagger_c\cdot\vec{A}\psi_c)+{{q\hbar}\over{2mc}}\psi^\dagger_c\vec{\sigma}\cdot\vec{B}\psi_c+q\psi^\dagger_c\phi\psi_c. \eqno{(20)}$$ 

To pass from (17) to (20), the identity $M^\dagger M=1$ is inserted at each term of (17), and the fact that $M\vec{\sigma}M^\dagger=M(\sigma_1,\sigma_2,\sigma_3)M^\dagger=(-\sigma_1,\sigma_2,-\sigma_3)$ is used; then the complex conjugation operation $K$ completes the transformation. 

\

The total action for the particle-antiparticle system is $$S_{tot}=S+S_c=\int dt \int d^3 \vec{x}({\cal L}(\vec{x},t)+{\cal L}_c(\vec{x},t)) \eqno{(21)}$$ and equation (5) is obtained from $S_{tot}$ or $S_c$ as $${{\delta}\over {\delta \psi^\dagger_c(\vec{x},t)}}S_{tot}={\delta \over {\delta \psi^\dagger_c(\vec{x},t)}}S_c=0. \eqno{(22)}$$

\

The lagrangian ${\cal L}$, and therefore the equation (1), are invariant under the galilean transformations (13), (15) and (16) for $\psi$, $\psi_c$, and $(\phi, \vec{A})$ and $\vec{B}$, respectively. To prove it, we use the facts that $\nabla=R^{-1}\nabla^\prime$ where $\nabla={{\partial}\over{\partial \vec{x}}}$ and $\nabla^\prime={{\partial}\over {\partial\vec{x}^\prime}}$, and ${{\partial} \over{\partial t}}={{\partial} \over{\partial t^\prime}}-R^{-1}\vec{V}\cdot\nabla^\prime$. If ${\cal L}^\prime$ and ${\cal L}^\prime_c$ are the transformed lagrangian densities for particles and antiparticles, then, from equation (20), $${\cal L}_c(\vec{x},t)=K{\cal L}(\vec{x},t)=K{\cal L}^\prime(\vec{x}^\prime,t^\prime)={\cal L}_c^\prime(\vec{x}^\prime,t^\prime) \eqno{(23)}$$ and therefore the galilean invariance of equation (5) is also proved. 

\

Finally, both ${\cal L}$ and ${\cal L}_c$, and therefore the equations (1) and (5), are gauge invariant under the transformations $\psi\to e^{i\Lambda}\psi$, $\psi_c\to e^{-i\Lambda}\psi_c$, $\phi \to \phi-{{\hbar}\over{q}}{{\partial}\over{\partial t}}\Lambda$ and $\vec{A}\to \vec{A}+{{\hbar c}\over{q}}\nabla\Lambda$, where $\Lambda$ is an arbitrary differentiable function of $(\vec{x},t)$.

\

{\bf Acknowledgement}

\

The author was partially supported by the project PAPIIT IN103505, DGAPA-UNAM, M\'exico. 

\

{\bf References}

\

1. A. Cabo, D. B. Cervantes, H. P\'erez Rojas and M. Socolovsky, Remark on charge conjugation in the nonrelativistic limit, {\it International Journal of Theoretical Physics} (to appear); arXiv: hep-th/0504223.

\

2. J. D. Bjorken, and S. D. Drell, {\it Relativistic Quantum Mechanics}, Mc Graw-Hill, New York (1964): p. 11.

\

3. M. Socolovsky, The CPT Group of the Dirac Field, {\it International Journal of Theoretical Physics}  {\bf 43}, 1941-1967 (2004); arXiv: math-ph/0404038.

\

4. V. B. Berestetskii, E. M. Lifshitz, and L. P. Pitaevskii, {\it Quantum Electrodynamics, Landau and Lifshitz Course of Theoretical Physics, Vol. 4}, 2nd. edition, Pergamon Press, Oxford (1982): p. 45.

\

5. J. A. de Azc\'arraga and J. M. Izquierdo, {\it Lie groups, Lie algebras, cohomology and some applications in physics}, Cambridge University Press, Cambridge (1995): p. 155.

\

6. S. Sternberg, {\it Group Theory and Physics}, Cambridge University Press, Cambridge (1994): p. 49.

\

\

\

\

\

e-mail: socolovs@nucleares.unam.mx, somi@uv.es

\

\end